\documentclass[12pt,preprint]{aastex}

\slugcomment{}
\shorttitle{The Central Star of the Planetary Nebula NGC~6302}
\shortauthors{}

\begin{document}

\title{Detection of the Central Star of the Planetary Nebula NGC~6302}

\author{C. Szyszka$^{1,2}$, J. R. Walsh$^3$, Albert A. Zijlstra$^2$, Y. G. Tsamis$^{4,5}$} 
\email{cszyszka@eso.org}

\altaffiltext{1}{European Southern Observatory, Karl-Schwarzschild Strasse 2, D-85748 Garching, Germany}
\altaffiltext{2}{Jodrell Bank Centre for Astrophysics, School of Physics \&\
  Astronomy, University of Manchester, Oxford Road, Manchester M13 9PL, UK}
\altaffiltext{3}{Space Telescope European Coordinating Facility, European Southern Observatory, Karl-Schwarzschild Strasse 2, D-85748 Garching, Germany}
\altaffiltext{4}{Instituto de Astrofis{\'i}ca de Andaluc{\'i}oa (CSIC), 
Apartado 3004, 18080 Granada, Spain}
\altaffiltext{5}{Department of Physics and Astronomy, The Open University, Walton Hall, Milton Keynes MK7 6AA, UK}

\begin{abstract}
NGC~6302 is one of the highest ionization planetary nebulae known and shows
emission from species with ionization potential $>$300eV. The temperature of
the central star must be $>200\,000$K to photoionize the nebula, and has been
suggested to be up to $\sim 400\,000$K. On account of the dense dust and molecular
disc, the central star has not convincingly been directly imaged until now. NGC~6302 was
imaged in six narrow band filters by Wide Field Camera 3 on HST as part of the
Servicing Mission 4 Early Release Observations. The central star is directly
detected for the first time, and is situated at the nebula centre on 
the foreground side of the tilted equatorial disc. The
magnitudes of the central star have been reliably measured in two filters
(F469N and F673N). Assuming a hot black body, the reddening has been measured
from the (4688-6766\AA) colour and a value of c=3.1, A$_{v}$=6.6 mag determined. A
G-K main sequence binary companion can be excluded.  The position
of the star on the HR diagram suggests a fairly massive PN central star of about
0.64\,M$_\odot$ close to the white dwarf cooling track. A fit to the 
evolutionary tracks for ($T,L,t$)\,=\,($200\,000$K, $2000$L$_\odot$, $2200$yr), 
where $t$ is the nebular age, is obtained; however the luminosity and 
temperature remain uncertain.  The model tracks predict that
the star is rapidly evolving, and fading at a rate of almost 1\%\ per
year. Future observations could test this prediction.
\end{abstract}

\keywords{planetary nebulae: individual (NGC~6302) --- stars: AGB and post-AGB --- stars: imaging}

\section{Introduction}
Planetary nebulae (PN) trace one of the fastest phases of stellar evolution. They
form when an Asymptotic Giant Branch (AGB) star ejects its envelope in a
massive wind, leaving only the remnant degenerate C/O core. After the
ejection, the star heats up rapidly, reaching temperatures over $100\,000$K,
within typically a few thousand years. The hydrogen burning ceases at this
point, and the star quickly fades by a large factor, before entering the white
dwarf cooling track. The hot star ionizes the previously ejected nebula: the
bright ionized nebula combined with the very hot, optically faint star, can
make the stellar emission difficult to detect. This is made worse if
circumstellar extinction hides the star from direct view in the optical and 
ultra-violet domain.

The central star of NGC~6302 has proven to be one of the most elusive of all
PN central stars. In spite of many attempts, it has not been
detected so far. NGC~6302 shows a strongly bipolar morphology, with
filamentary emission extending over 7$'$.  Bipolar morphologies are
thought to be associated with circumstellar discs and torii \citep{Balick1987, 
Icke2003}, which are commonly seen in post-AGB nebulae \citep{Siodmiak2008}.
NGC~6302 provides an extreme case for such a massive, optically opaque torus
\citep{Matsuura2005, Peretto2007,Dinh2008}.

The central star of NGC~6302 is believed to be extremely hot, as indicated by
the optical spectrum \citep[e.g.][]{Tsamis2003} and observation of infrared
coronal lines of highly ionized elements, of which the highest observed is
Si$^{+8}$ with an ionization potential of 303eV \citep{Casassus2000}. To
produce such high ionization species various authors have suggested very hot
stars: \citet{Ashley1988} proposed a black body of 430\,000 K;
\citet{Pottasch1996} $380\,000$K; and \citet{Casassus2000} 
$250\,000$K.  \citet{Groves2002}, based on the observed scattered continuum,
estimated the temperature of the central source to be $150\,000$K. Most
recently, \citet{Wright2007}, based on 3D photo-ionization modeling, proposed
a hydrogen deficient central star with a surface temperature of $220\,000$K.

These stellar temperatures are similar to those of NGC~7027, another very bright,
compact PN at a similar distance. But whilst the star of NGC~7027 has been
unambiguously identified in images \citep[e.g.][]{Wolff2000}, the star of
NGC~6302 has remained hidden even in previous HST and ground-based images.
The only possible detection is from \citet{Matsuura2005}, who located a
compact source emitting at 3.94 $\mu$m, possibly associated with the [Si~IX]
emission line.  It is not known whether the star of NGC~6302 is fundamentally
different from NGC~7027 (higher temperature or lower luminosity). 
We here present new HST observations with the newly-installed WFC3 instrument, 
which for the first time directly reveal the
central star, and allow us to address these questions.

\section{Observations}
NGC~6302 was observed as part of the HST Early Release Observations (ERO) to
demonstrate the scientific capabilities of the newly installed Wide Field
Camera 3 (WFC3) on board the Hubble Space Telescope (HST). At optical
wavelengths, the UVIS channel has two 2k$\times$4k CCDs, with a spatial scale of 39
mas/pixel and a field of view of 162 arcsec.  Imaging programme 11504 (PI:
K. Noll) was executed on 2009 July 27 using 9 orbits.  Table~1 lists the
observations in six narrow band filters that isolate emission lines from
ionization potentials of 10eV ([S~II] 6716 \& 6731 \AA) to $>$54eV (He~II 4686\AA).

\section{Reduction and Results}
The pipeline-reduced long-exposure images for all the filter bands (i.e. FLT
files) were combined using {\em multidrizzle} \citep{Fruchter2009} rebinning
the output to 0.040$''$/pixel.  The images in different filters were matched
using field stars in common. The alignment was verified by inspecting star
positions in the final images.  Comparison of the multidrizzled images with
those in the public ERO release showed good agreement.

Figure 1 shows the F673N filter image of the central arcminute of NGC~6302 
and the 2.4$''\times$2.4$''$ area around the
central star, for all the six HST WFC3 filter images 
in the ERO release.  This star is well detected on the F469N and F673N
filters at the position $\alpha = 17^h13^m44.39^s,\delta = -37^{\circ}06'12.''93$ 
(J2000) and marginally detected on the F502N, F656N and F658N images.  From
the filter parameters summarized in Table 1 and the emission line spectrum of
NGC~6302, as taken for example from \citet{Groves2002}, the only emission
line contribution to the F469N filter is He~II 4686\AA. The F673N filter
contains weak lines of He~I and C~IV, in addition to the [S~II]6730\AA\ line;
the [S~II]6716\AA\ line is just at the half power point. All the other filters
contain strong emission lines. The ease of detection of the
star in the F673N filter is accounted for by the lower emission line
contribution to the band, producing a much higher star/background
contrast. This also applies to the F469N filter, where the He~II line is 
$^{<}_{\sim}$10\% of the strength of the [O~III]5007\AA\ line (in the
F502N filter) and of the H$\alpha$ line in F656N.

Photometry was then performed on the star at the centre of NGC~6302 in the six
combined multidrizzled images. Robustness of the results depends on accurate
determination of the high and structured background. We applied three
different approaches. First, using a small aperture (2 pixels radius) and a 3
pixel wide 'sky' (in reality nebula) annulus.  Second, a surface fit to the 
local background, excluding a small
annulus around the central star, was made in the same way for all images:
this was subtracted from the image, and aperture photometry was performed on
these images. As a third approach, Sextractor \citep{Bertin1996} photometry
was performed. For the F469N and F4673N filter images the results showed good 
consistency, but for the other images there is a larger scatter of typically
$\pm$0.3mag. In Table 1, the 2.0 pixel radius photometry is listed, where an
aperture correction was applied by performing photometry of uncrowded stars in
the periphery of the nebula with increasing aperture size.

Figure 2 shows the photometric points (in AB magnitudes) plotted as a function
of the wavelength. At the position of each point, a horizontal bar shows the
width of the respective filter.  Also plotted on the X-axis (in green) is a
compressed log spectrum of NGC~6302 taken from \citet{Groves2002}, applying a 
logarithmic extinction at H$\beta$ of $c = 1.2$. A nebular 
continuum (computed for $T_e=15000$K and He/H\,=\,0.18) was added 
to this emission line spectrum and the result converted to WFC3
filter photometry using the STSDAS.synphot package. This allows comparison
of the stellar photometry with the colours expected from the nebula spectrum, 
the latter shown as red diamonds on Figure 2 (with arbitrary absolute flux scaling tied 
to the F656N filter measurement of the stellar magnitude). From this
comparison it is clear that the photometric points for the F469N and F673N
filters are inconsistent with nebular
emission. Although there appears to be a feature in the F373N, F502N, F656N
and possibly F658N, filters, at the position of the star in the F469N and
F673N images (see Figure 1), the heavy nebula contribution prevented 
reliable measurement of the stellar magnitudes and the results correspond
more closely to that of (inadequately subtracted) nebula emission.  The
stellar magnitude is marginally measurable on the F502N image. Even though the
stellar flux is higher in the F373N band, the high extinction prevents
reliable detection in this filter.

Taking the two photometric points at 4690 and 6766\AA\ (Table 1) as reliable
detections of a putative central star of NGC~6302, the (4690--6766\AA) colour
was compared with that of a star of a given black body temperature subjected
to line-of-sight extinction. Given the very high temperature estimates for the
central star, a very large value of extinction must be applied to produce the
observed (4690--6766\AA) colour of 2.1 mag. Two extinction laws were used to
determine the extinction - the standard Galactic law of \citet{Seaton1979}
with the ratio of total to selective extinction of $R = 3.2$ and, on the basis
of the good agreement of the observed and theoretical nebula continuum
determined by \citet{Groves2002}, the Whitford law \citep[as tabulated 
by][]{Kaler1976a}, for the three estimates of the stellar temperature, using
a black body. For such high temperatures, a black body should be a good
approximation to the optical colours. Table 2 lists the values of c determined
in these three cases. The values of $c$ can be directly converted to $A_V$ by 
multiplying by 2.18 and are listed in Table 2.

For such high temperatures there is not sufficient sensitivity to determine
the temperature of the central star, and the extinction only changes by 0.02
for black body temperatures from $150\,000$ to $400\,000$K. From the unreddened flux,
taking the F673N magnitude as the most reliable, assuming a black body, and the
distance of 1.17\,kpc \citep{Meaburn2008}, the derived stellar luminosity can be
determined and is listed in Table 2.

\section{Discussion}
The HST point spread function and the excellent optical and CCD quality
of WFC3 UVIS have allowed the central star of NGC~6302 to be
unambiguously detected for the first time.  The star is situated in the
centre of the large-scale nebular outflows, as expected, and lies on the
eastern edge of the thick dust lane. This location is also the expected one
given that the eastern lobe is tilted to the foreground \citep{Meaburn2008}.
However the central star is not at the expected position at the centre of the 
innermost torus as shown by the radio data \citep{Matsuura2005}. No optical 
emission counterpart was found at the position of the infrared source of 
\citet{Matsuura2005}, which is approximately $2.''6$ north from the proposed 
central star. Nor was a point source detected at the position of the 6cm 
radio continuum emission peak reported by \citet{Matsuura2005} (about $2.''2$ 
from the star); only an upper limit of 23.8 (AB mag.) was measured for the F673N 
magnitude of a source at this position by aperture photometry. The 
(4688-6766\AA) colour at this radio peak position resembles nebular emission. 
Thus if a star was present at this position, it would suffer much higher extinction 
than indicated by the \citet{Matsuura2005} extinction maps. 

The very high extinction towards the central star derives mostly from the
circumstellar torus. From the ratio map of H$\alpha$ to radio flux,
\citet{Matsuura2005} derive $A_{H\alpha}= $ 5--7 for the central region,
corresponding to $A_V = $ 6--8.  Field stars indicate a foreground
extinction $A_V=1.1$ \citep{Matsuura2005}, leaving the remainder as
circumstellar extinction.  The Br$\alpha$ image of \citet{Matsuura2005} shows
that the peak extinction is considerably higher than derived from H$\alpha$,
possibly  as high as
$A_V=30$: H$\alpha$ cannot be detected from such regions. The region of highest
extinction is offset from the centre, consistent with the fact that the torus
is seen with a small inclination of 18$^\circ$ \citep{Meaburn2008, Peretto2007, 
Dinh2008}.

The distance to NGC~6302 determined by \citet{Meaburn2008} as 1.17\,kpc from
expansion proper motions of filaments agrees
fairly well with an assumed distance of 0.91\,kpc from \citet{Kemper2002} and
1.0\,kpc from \citet{Matsuura2005}. Older values include \citet{Rodriguez1985} 
who argue for a distance of
1.7\,kpc, with a possible association to the star formation region NGC 6334.
Distances much larger than 1\,kpc lead to very high masses for the shell and
torus \citep[e.g.][]{Matsuura2005}.

The nebular luminosity, scaled to a distance of 1.17\,kpc, was determined as
$3.3 \times 10^3 \,\rm L_\odot$, based on infrared dust modeling
\citep{Matsuura2005}.  An almost identical value is derived from the radio
flux of \citet{Rodriguez1985}. Both measure the reprocessed stellar
radiation, and, for an optically thick nebula, the nebula luminosity should
track the luminosity of the star. To derive a stellar luminosity from the
photometry, a temperature and a spectrum must be assumed. For a black body,
Table 2 shows that the stellar and nebular luminosity agree well for a stellar
temperature of around $200\,000$K.

Many central stars of PNe are expected to be binaries \citep{deMarco2009}. The
central star of NGC 6302 has also been suggested to be binary:
\citet{Feibelman2001} claimed evidence for a G5V companion based on IUE
spectra, although this has been disputed by \citet{Meaburn2005}. A G5V star 
has $M_V = +5.1$, and for a distance of 1.17kpc and an extinction of $A_v=6.7$, 
the observed V magnitude on the Vega system would be V=22.1. Interpolating the 
observed magnitudes in Fig. 2, the AB mag at V (very close to the Vega system) 
is 21.5, not incompatible with a G5V star. However, for a temperature of 5200K, 
suggested by Feibelman, with an extinction of 6.7 mag, the (4688--6766\AA) colour would be 
1.5 mag. larger than observed.  A much lower extinction ($c=1.4$) could reproduce the
(469--673) colour, but the absolute V mag would be around 8.1. corresponding to
a K5 star. Thus, we can exclude a G5V star. 
A K5V star is
in principle possible, but requires a much lower extinction than measured from
H$\alpha$. In contrast, a hot star precisely fits the measured extinction and
is the preferred interpretation. If we were detecting the companion, the true
central star would be hotter and/or less luminous than derived here (Table 2).

Knowing an accurate luminosity and temperature allows one in principle to
obtain the mass of the star, an important parameter for the stellar modeling.
At this phase of evolution, the stellar mass is close to the mass of the
degenerate C/O core, or the later white dwarf. Typically, this method does not
provide accurate masses. The luminosity is typically too uncertain to allow
for an accurate mass determination, because of distance errors. In the case of
NGC 6302, the luminosity is accurate to 25 per cent \citep[based on a distance 
error of 12\% from][]{Meaburn2008}, however the star appears to be on
the cooling track where  tracks of different masses are closely clustered.

Instead, the evolutionary time scale can be used to improve on the mass
accuracy: the speed of the evolution along the \citet{Bloecker1995}
track is a very strong function of stellar mass.  The surrounding nebula acts as
a clock, as its expansion dates from the time the star departed the AGB. This
age, plus the stellar temperature and luminosity, can be used to derive an
accurate stellar mass. This method is described in \citet{Gesicki2007} and
references therein. The age of the nebula can be derived from the angular extent, distance, and
de-projected expansion velocity. The CO observations of \citet{Peretto2007}
show that the torus ejection ended 2900\,yr ago. The bipolar lobes appear to
be a little younger, with an age of 2200\,yr \citep{Meaburn2008}. 
We will assume that 2200 yr represents the full post-AGB evolution age of the
central star.

A luminosity of 4000\,L$_\odot$ puts the star at the upper end of the cooling
track. We use interpolations of the Bloecker tracks to obtain a grid of models
at a mass resolution of 0.02\,M$_\odot$, as described by \citep{Frankowski2003}.  
The 0.68M$_\odot$ track passes through
the point at ($T=220\,000$K, L=4000L$_\odot$), however it does so at a post-AGB
age of  1000\,yr which is too young to fit NGC 6302. Higher mass tracks are hotter at this
luminosity, and much younger, while lower mass stars are older and a little
cooler. Again, the tracks run very close in the HR diagram, leaving the
age as the important discriminant. Figure 3 shows the possible locations of 
the star (for both extinction laws in
Table 2), the interpolated Bloecker tracks and the locus of tracks of
different masses for an age of 2200 yr. The lines intersect at
($T,L$)\,=\,($200\,000$K, $2000$L$_\odot$), corresponding to a stellar mass of
$M = 0.646\,\rm M_\odot$. This fit gives a temperature a little below that
derived by \citet{Wright2007}. The different extinction laws give essentially
the same mass: this is because the evolution at this location in the HR
diagram is very fast.

It should be noted that the fit is only as accurate as the models, and there
is a considerable systematic uncertainty in the derived mass
\citep{Zijlstra2008}. The Bloecker tracks are calculated for a sparse range of
masses only, so that the interpolation is done over a relatively large mass
range. The speed of evolution especially of the early post-AGB evolution, is
uncertain, and this uncertainty feeds through to the time scales at later
stages. However, the method is expected to give good constraints on relative
masses. Thus, comparison with the results of \citet{Gesicki2007} shows that
NGC 6302 has a high mass compared to typical PNe.

The result can also be compared to that of another high-mass PN, NGC~7027.
\citet{Zijlstra2008} derive a mass of $0.655\pm0.01$M$_\odot$ using the same
method. The star of NGC 7027 is still on the horizontal, high luminosity
track, and the nebula is much younger. For an age of 2200\,yr, the star of
NGC~7027 will have evolved much farther down the cooling track: the
interpolated Bloecker tracks predict $L=250\,\rm L_\odot$. The comparison
therefore suggests that the star of NGC~6302 is indeed a little less massive
than that of NGC~7027.

The fit to the star of NGC~6302 suggests it is currently in a phase of
very fast evolution. The model track shows a decline of the luminosity
of $\sim 0.8$ per cent per year.  Such fast evolution could leave the
ionization of the nebula out of equilibrium. 

In summary, this paper presents the first direct detection of the central star
of NGC 6302, among the hottest stars of planetary nebulae. The photometry
indicates a hot, significantly extincted star. The location in the HR diagram and
age of the nebula indicates a massive star, with mass of about
0.64\,M$_\odot$. The models suggest that a star at this location would
be in a phase of rapid fading, and it would be of interest to test this
observationally.

\acknowledgments{This paper uses data obtained with \facility{HST (WFC3)}.
YGT acknowledges support from grants AYA2007-67965-C03-02/CSD2006-00070/CONSOLIDER-2010 
of the Spanish Ministry of Science. Also CS gratefully acknowledges an ESO 
studentship and JBCA bursary.}

\bibliographystyle{apj}

\begin{table}
\begin{center}
\caption{Exposure details, basic properties of filters used and measured AB and Vega magnitudes of the central star.\label{tbl-1}}
\begin{tabular}{ccccccccc}
\tableline\tableline
Filter   	 &$\lambda$    	& FWHM 		&Emission 	&\multicolumn{2}{c}{Exposures} &AB    &Error   &Vega \\
         	 &[\AA]       	& [\AA]         &Line	&Long &Short &[mag] &[mag]   &[mag]         \\
\tableline
   F373N          &3730.2           &39.2	&[O II]     &4$\times$1400s &\nodata		&25.0    &0.30    &24.1 \\
   F469N          &4688.1           &37.2	&He II      &4$\times$1400s &\nodata   		&22.47   &0.10    &22.63 \\
   F502N          &5009.6           &57.8	&[O III]    &6$\times$370s  &1$\times$35s     	&21.6    &0.30    &21.7 \\
   F656N          &6561.4           &13.9	&H${\alpha}$&6$\times$350s  &1$\times$35s   	&19.3    &0.20    &18.7 \\
   F658N          &6584.0          &23.6	&[N II]     &6$\times$370s  &1$\times$35s     	&19.0    &0.20    &18.6 \\
   F673N          &6765.9           &100.5	&[S II]     &4$\times$1300s &1$\times$100s     &20.33   &0.07    &20.08 \\
\tableline
\end{tabular}
\end{center}
\end{table}

\begin{table}
\begin{center}
\caption{Derived extinction, dereddened flux and luminosity of the star
(assuming a black body), for three 
published temperatures. $A_V$ is given for the Seaton reddening curve (upper)
and Whitford reddening curve (lower), as tabulated by \citet{Kaler1976a}.
The 4th column lists the dereddened stellar continuum flux at
6766\AA (F673N).\label{tbl-2}}
\begin{tabular}{ccccc}
\tableline\tableline
$T_{\rm BB}$    &$c$	&$A_{\rm V}$ & F(6765\AA)    & $L$ \\ 
 \,[K]         &        & [mag]  & [erg\,cm$^{-2}$\,s$^{-1}$]        & [L$_\odot$] \\
\tableline
\multicolumn{5}{c}{Seaton law}\\
150000 & 3.05  & 6.65 & 1.84E-15   &   1280 \\
220000 & 3.06  & 6.67 & 1.86E-15   &   4010 \\
400000 & 3.07  & 6.69 & 1.89E-15   &  24000 \\
\tableline
\multicolumn{5}{c}{Whitford law}\\
\tableline
150000 & 2.88  & 6.28 & 1.23E-15   &   860 \\
220000 & 2.89  & 6.30 & 1.25E-15   &   2700 \\
400000 & 2.90  & 6.32 & 1.27E-15   &   16100 \\
\tableline
\end{tabular}
\end{center}
\end{table}

\begin{figure}
\begin{center}
\includegraphics[width=0.75\textwidth]{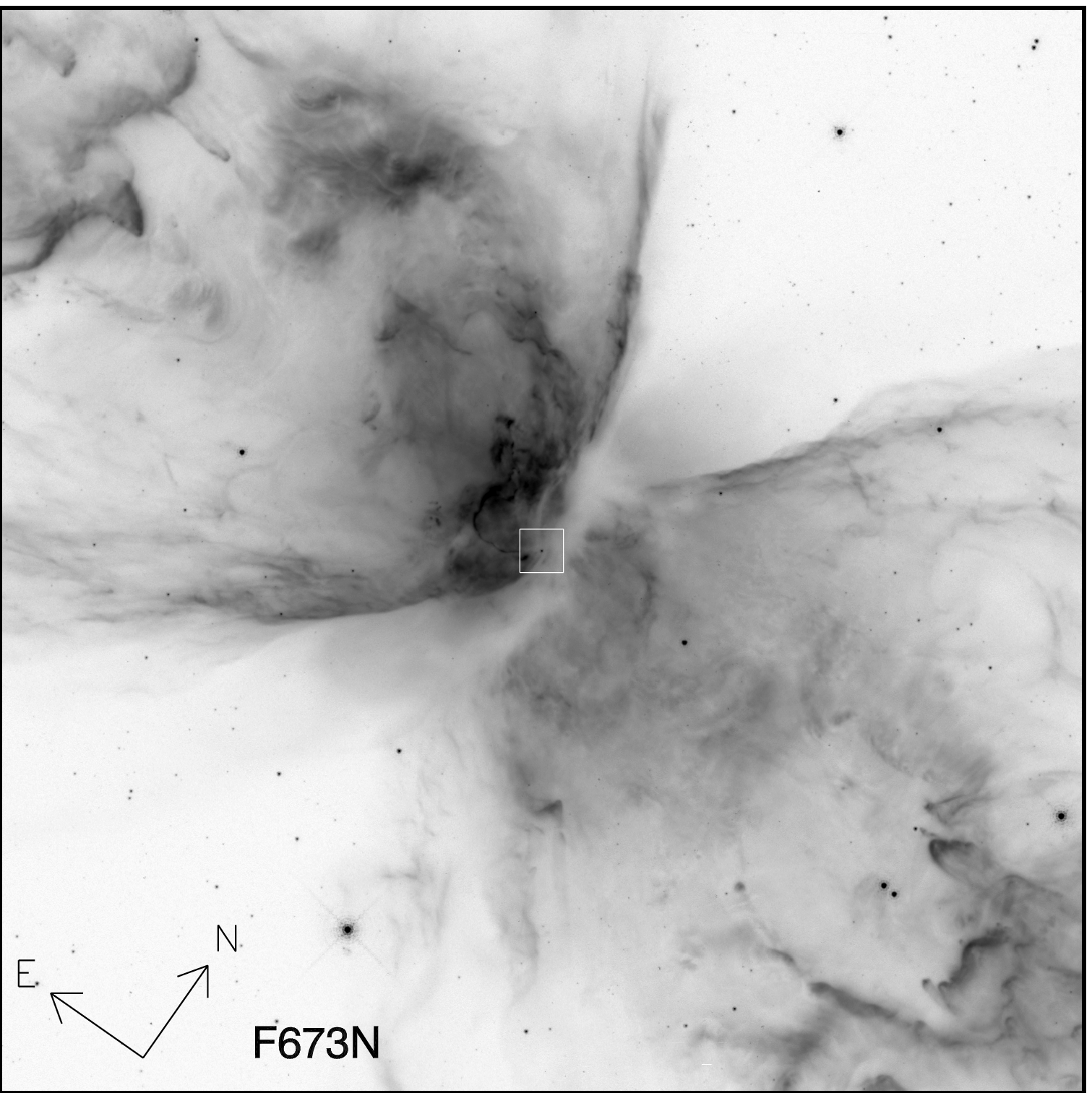}
\includegraphics[width=0.25\textwidth]{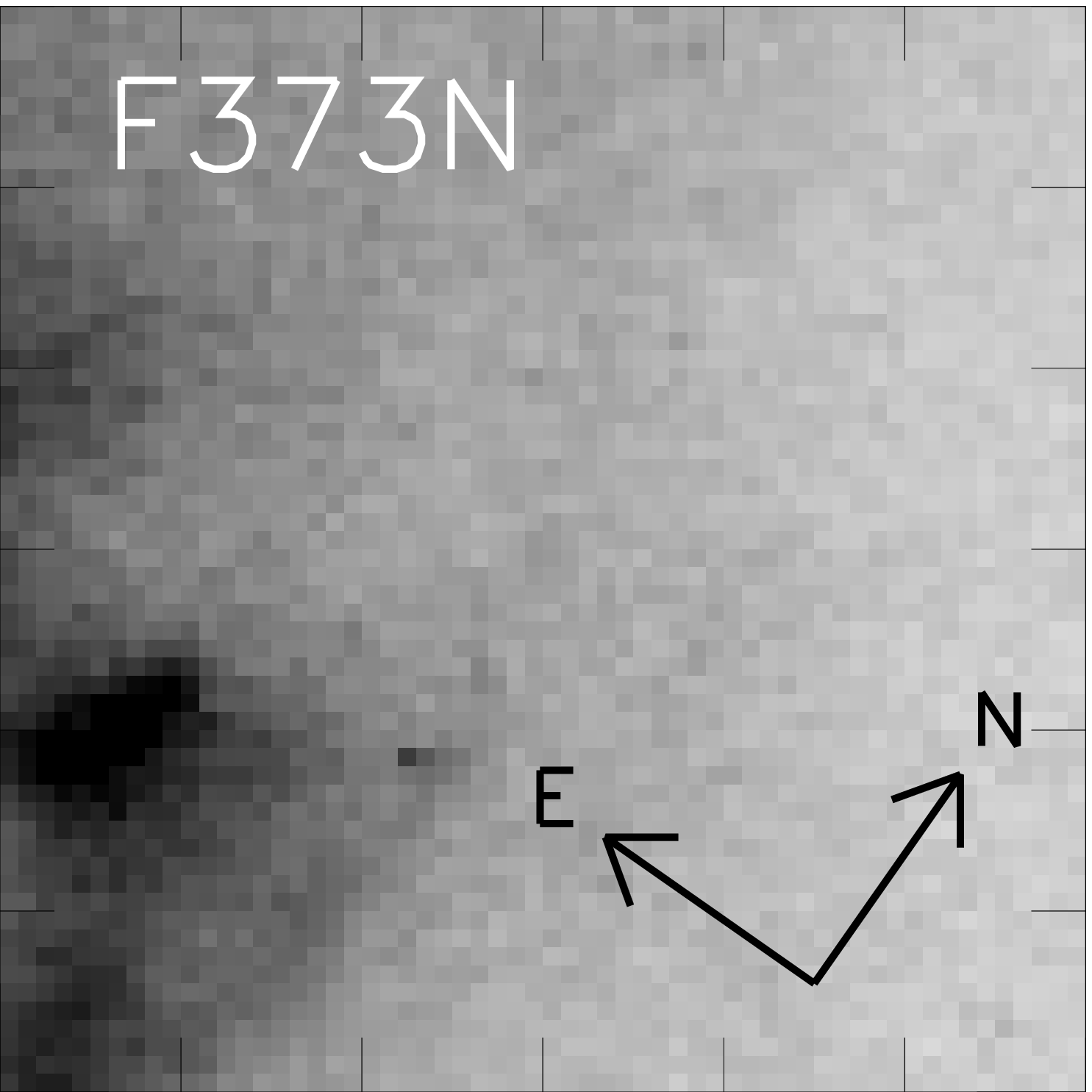}
\includegraphics[width=0.25\textwidth]{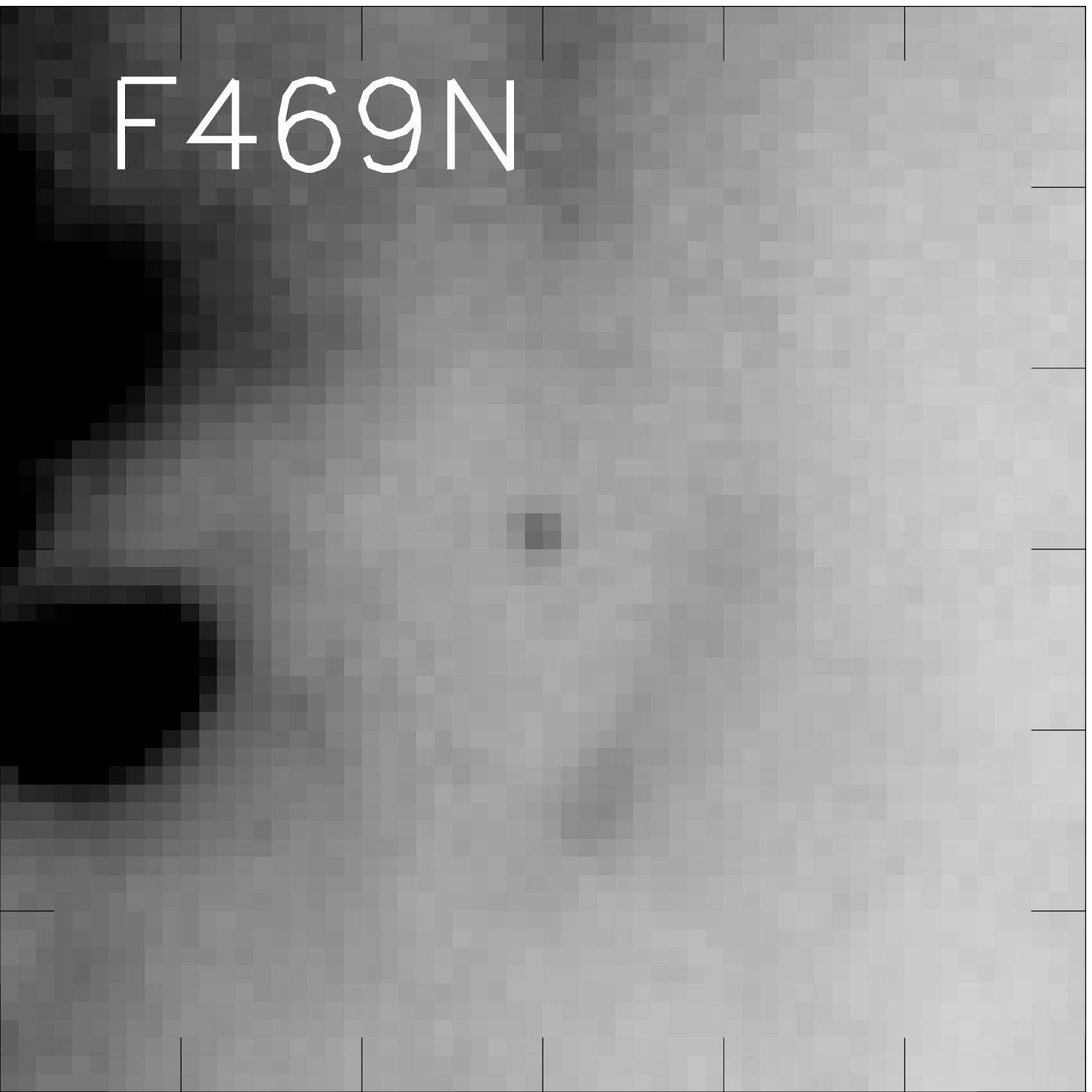}
\includegraphics[width=0.25\textwidth]{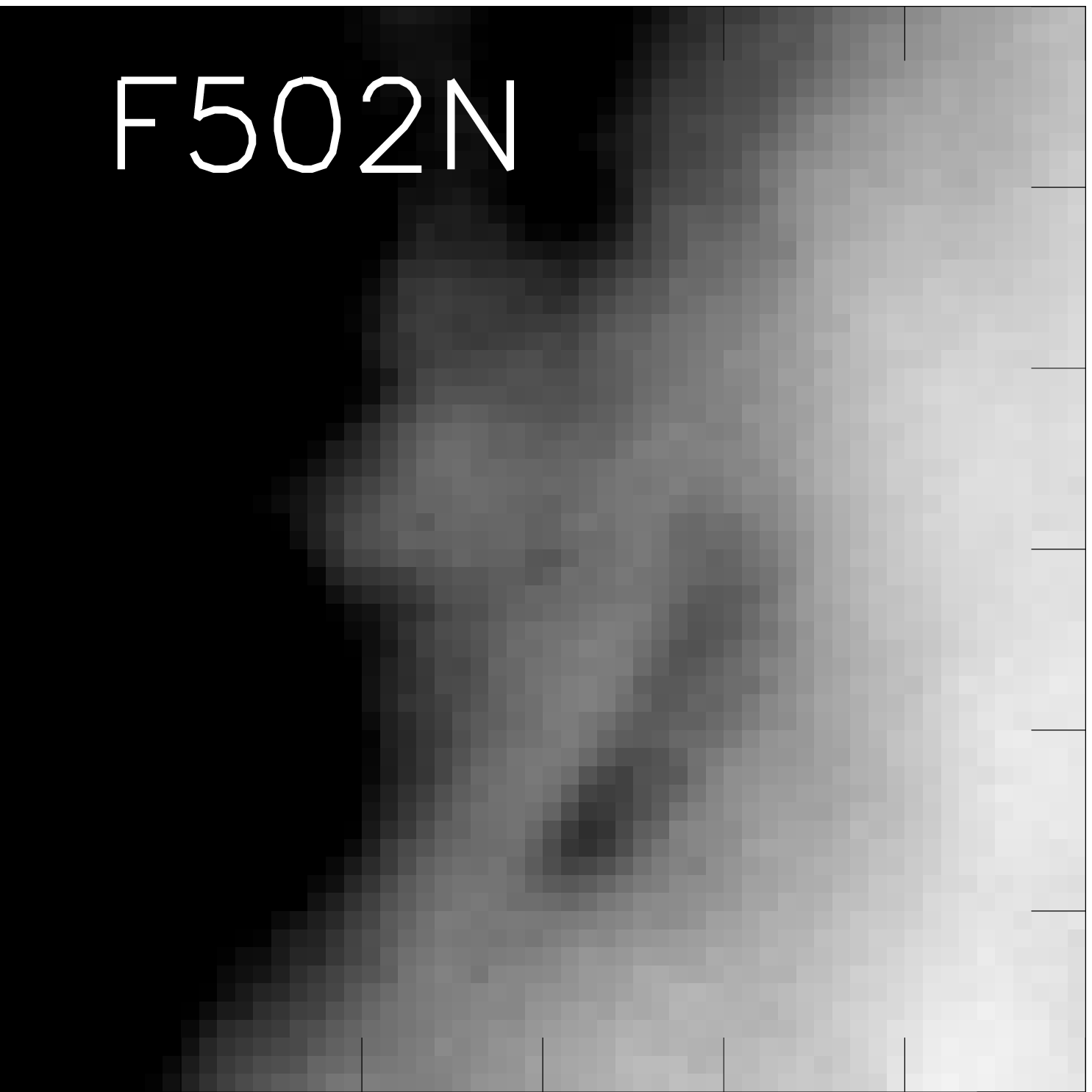}
\\
\includegraphics[width=0.25\textwidth]{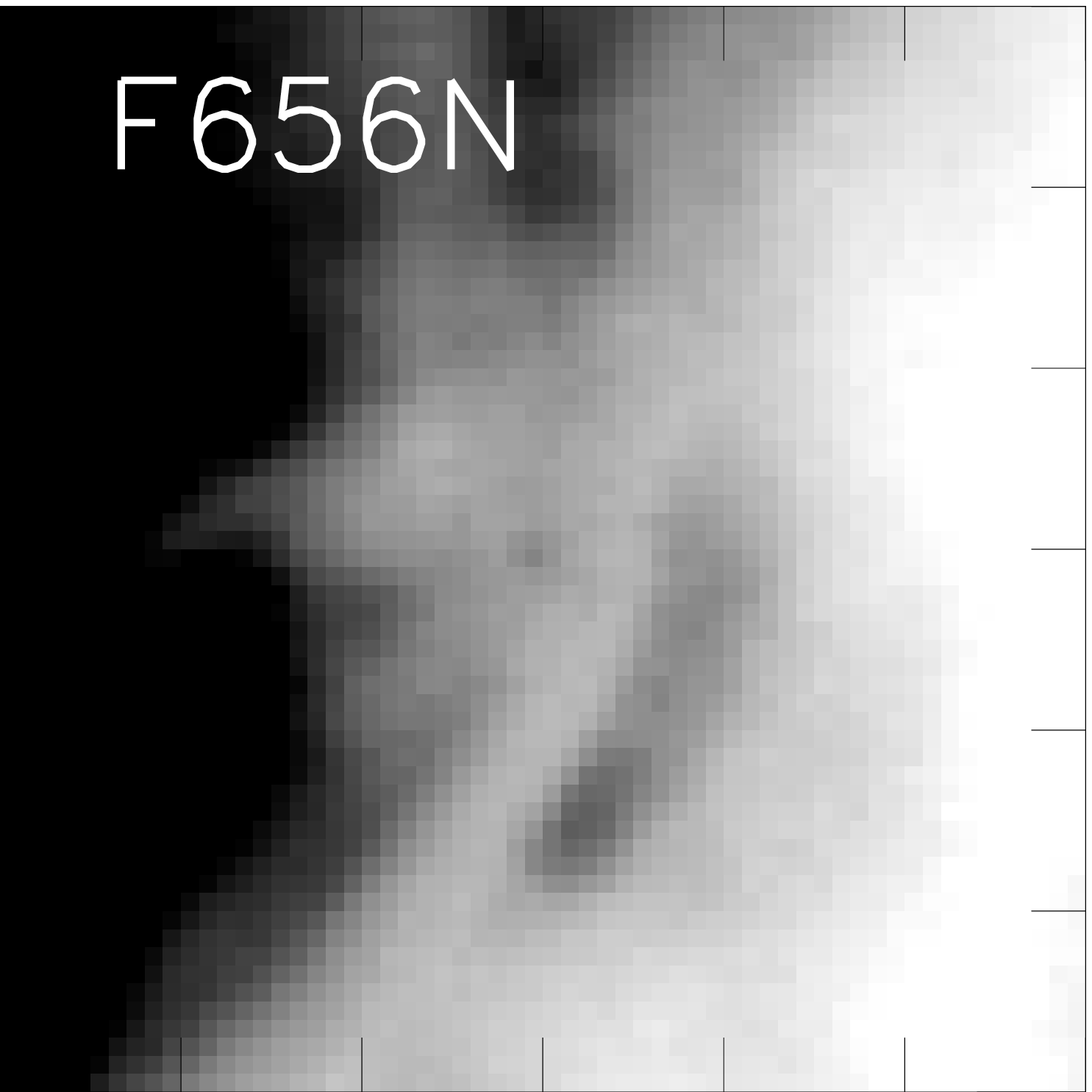}
\includegraphics[width=0.25\textwidth]{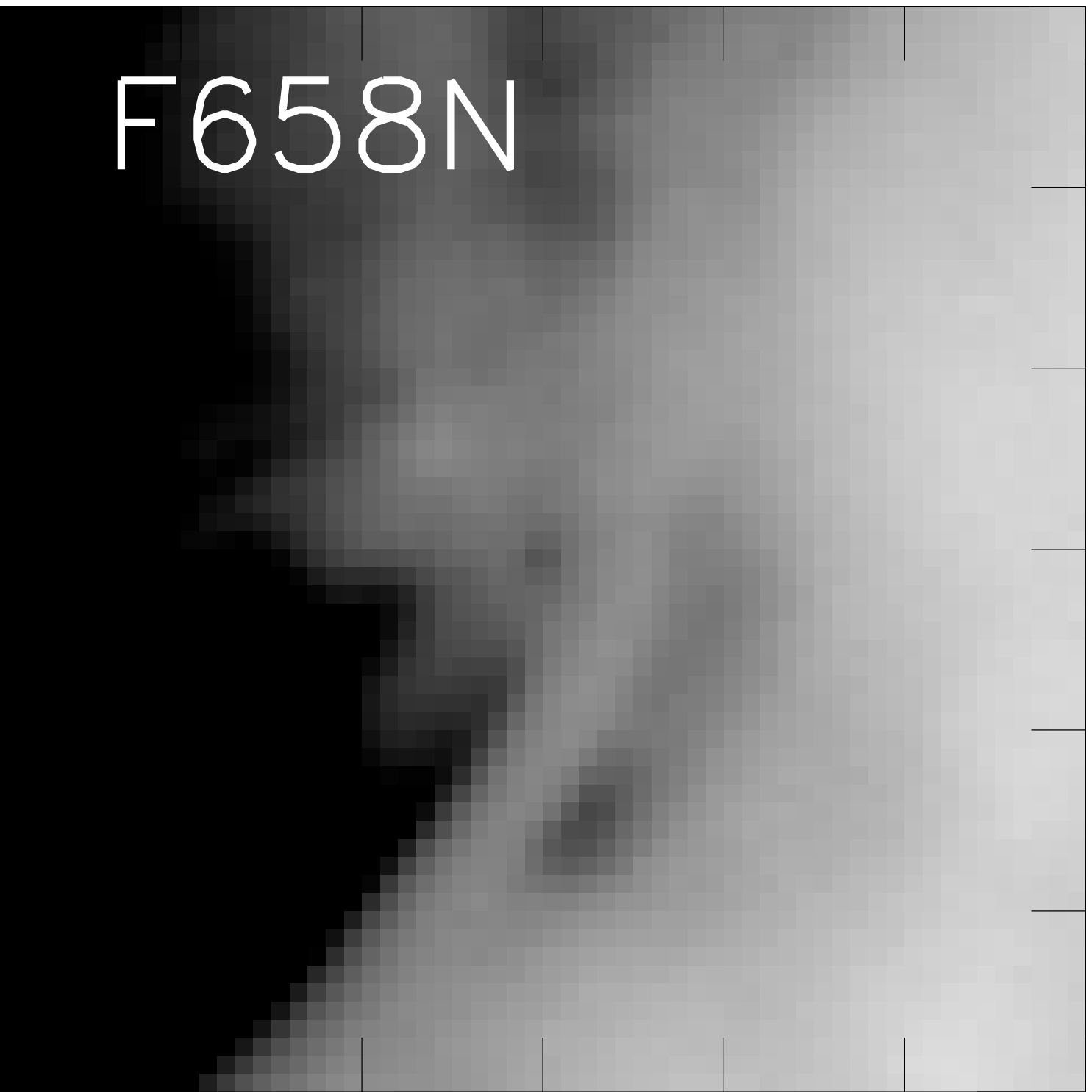}
\includegraphics[width=0.25\textwidth]{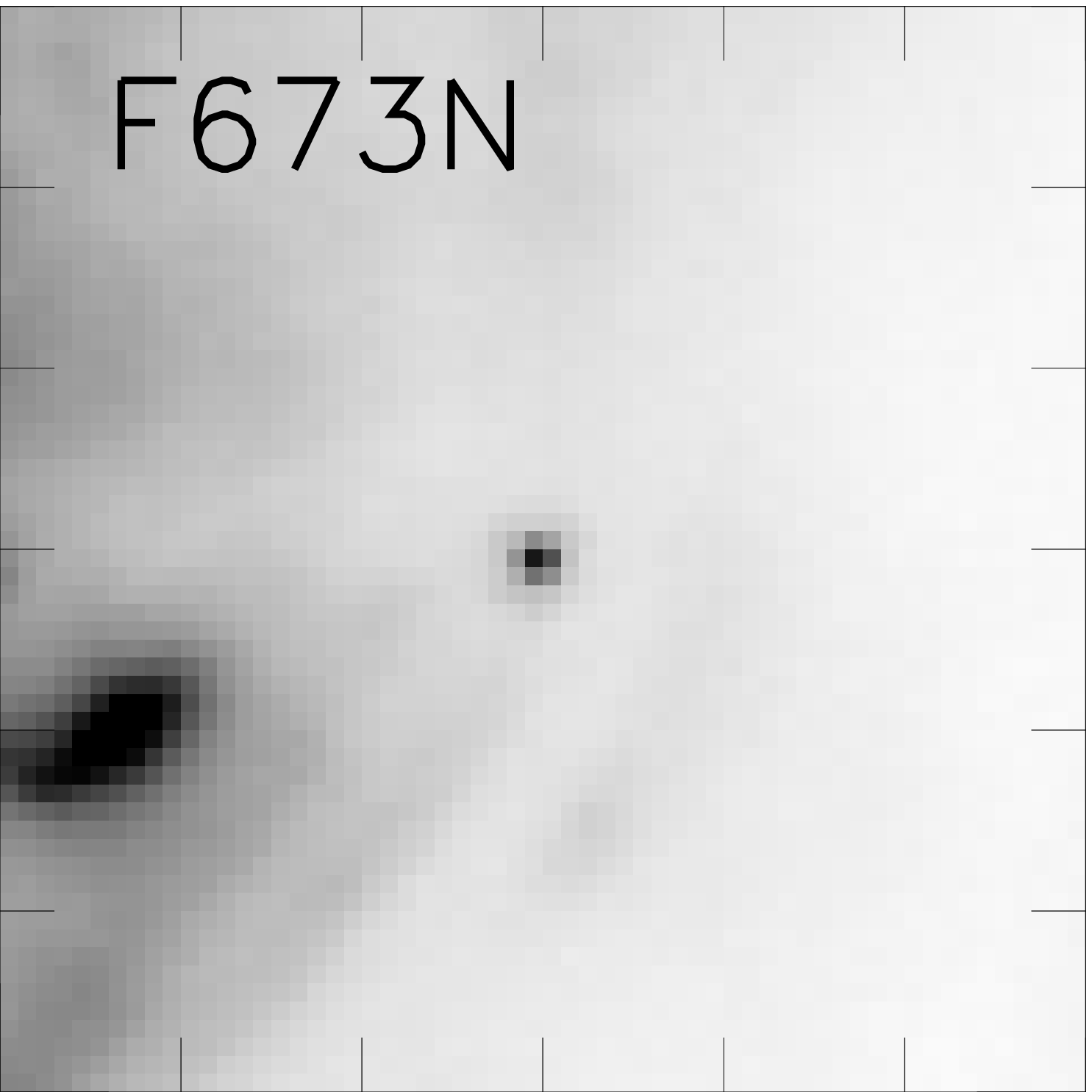}
\caption{Overview of the central $1'\times1'$ field of NGC~6302 in the
F673N filter (upper image). White square indicates 2.4$''\times$2.4$''$ field 
of six smaller images (below), with the central star in the middle of each image. 
Filter names are indicated and orientation. \label{fig-1}}
\end{center}
\end{figure}

\begin{figure}
\begin{center}
\includegraphics[width=0.8\textwidth, angle=270]{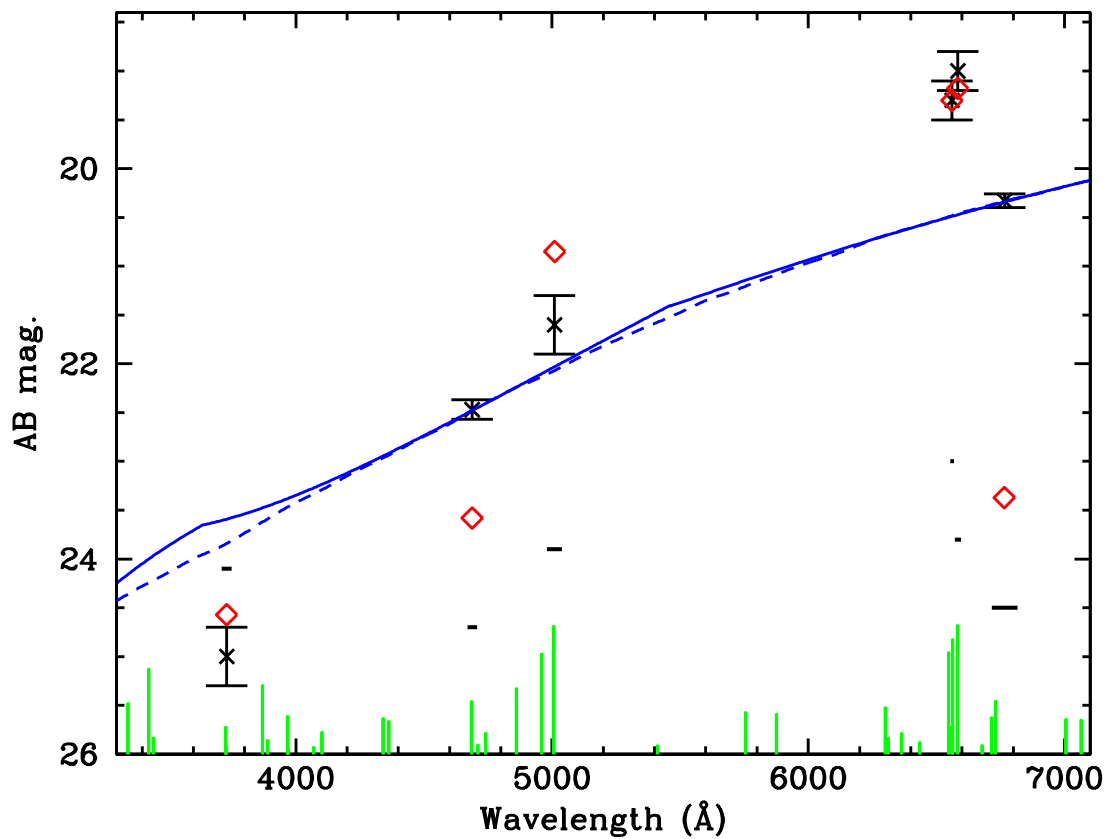}
\caption{The HST narrow band filter magnitudes (AB mag.) for the central star
of NGC~6302 are plotted as a function of the wavelength. At the position
of each filter passband a horizontal bar shows the width of the respective
filter. The full blue curve shows the fit of the photometry by a
$220\,000$K black body extinction corrected by a \citet{Seaton1979} Galactic
reddening of c=3.05, and the dashed blue curve the match with
a Whitford reddening law with c=2.89. (The two features in the Seaton reddening 
curve occur at the boundaries of the numerical fit functions to the extinction
law.) Also plotted (in green) is a
compressed log spectrum of NGC~6302 taken from Groves et al. (2002).
The red diamonds show the WFC3 filter magnitudes computed for the nebula,
using the emission line spectrum of Groves et al. (2002) with the
addition of nebular continuum, scaled to the F656N stellar magnitude. \label{fig-2}}
\end{center}
\end{figure}

\begin{figure}
\begin{center}
\includegraphics[width=0.8\textwidth]{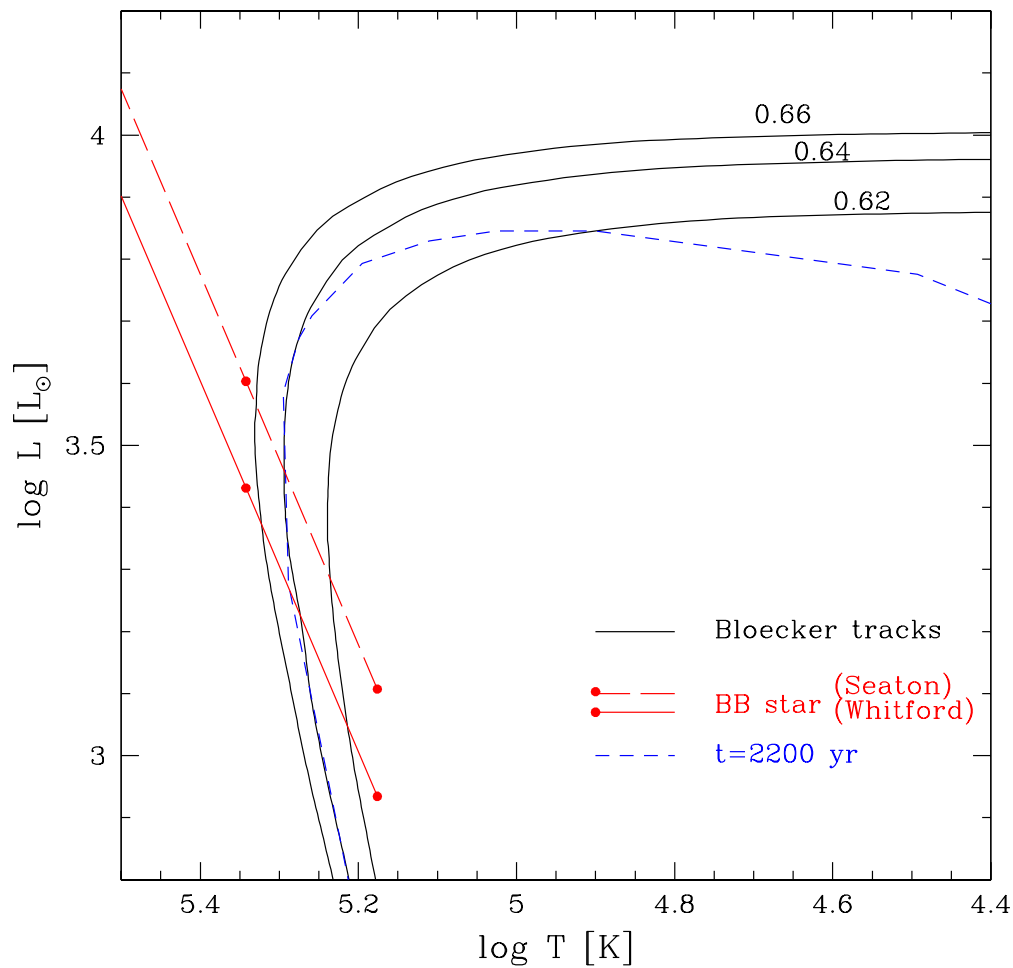}
\caption{The central star of NGC~6302 on the HR diagram.
  Red lines show the possible locations of the star according to values in Table 2:
  dashed line corresponding to dereddening with Seaton law, solid line for Whitford law. The
  black continuous lines show the interpolated Bloecker tracks, labeled with their
  mass (M$_\odot$). The blue dashed line shows the locus of all tracks
  for a post-AGB age of 2200 yr. The intersection suggests a mass 
  around 0.64\,M$_\odot$.
\label{fig-3}}
\end{center}
\end{figure}

\end{document}